\documentstyle[12pt]{article}
\input tcilatex
\begin{document}

\begin{center}
{\bf ON THE STABILITY OF SPHERICALLY SYMMETRIC CONFIGURATIONS IN NEWTONIAN
LIMIT OF JORDAN, BRANS-DICKE THEORY. }

\vspace*{1.5cm} S.M.KOZYREV

e-mail: Kozyrev@e-centr.ru
\end{center}

\vspace*{1.5cm}

\begin{center}
${\bf Abstract}$
\end{center}

We discuss stability of spherically symmetric static solutions in Newtonian
limit of Jordan, Brans-Dicke field equations. The behavior of the stable
equilibrium solutions for the spherically symmetric configurations
considered here, it emerges that the more compact a model is, the more
stable it is. Moreover, linear stability analysis shows the existence of
stable configurations for any polytropic index.

\section{Introduction}

Scalar-tensor theories, in which a long-range scalar field in addition to
the usual tensor fields present in Einstein's theory mediates gravity, have
been studied in many works (see e.g.\cite{1},\cite{10} ) as natural
generalizations of Einstein's general relativity (GR). The simplest of them
Jordan, Brans-Dicke (JBD) theory of gravity \cite{2},\cite{20}, in which a
scalar field $\phi $ acts as the source for the gravitational coupling with 
{\it G} $\sim \frac 1\phi $, was essentially motivated by apparent
discrepancies between observations and the weak-field predictions of GR. It
is a well-known fact that most of the mathematical difficulties of theories
of gravity lie in the high non-linearity of the field equations. However,
under the special circumstance when the gravitational field is weak one can
linearize the field equations thereby ignoring this feedback effect. Then
the weak-field limit is analyzed and the conditions leading to significant
deviations of the 1/r$^2$ Newton's law of gravitation are discussed.

The stability of equilibrium configurations for both stars and star clusters
remains a topic of continuing interest. The fundamental dynamics describing
galaxies, clusters of galaxies, or globular clusters \cite{3}, is a
collisionless system. In the general case the collisionless Boltzmann
equation cannot be solved because it involves too many independent
variables. However, we can get certain system with a polytropic state
equation, corresponding to isotropic velocity dispersion tensors. Moreover,
spherically symmetric static Newtonian perfect fluid models are the starting
point for many discussions about stellar structure and evolution. Within
this class of models, polytropic equations of state have been studied
thoroughly \cite{c31}, \cite{c32},\thinspace \cite{c33}, \cite{c34}, \cite
{c35}

\[
P=K\rho ^\gamma \equiv K\rho ^{1+\frac 1n,} 
\]

where K is non-negative polytropic constant, $\gamma $ is the adiabatic
index, and {\it n} the politropic index. In astrophysics pure polytropes
have been considered and some of the corresponding stellar models have been
studied in detail in many textbooks \cite{4}, although mostly numerically.

In this paper we consider static spherically symmetric perfect fluid models
using Newtonian approximations of Jordan, Brans-Dicke (JBD) theory of
gravity. This paper is organized as follows. In section 2 we have
constructed the field equations for the Newtonian limit of JBD theory. In
section 3 we derive the basic equation with approximation of small linear
perturbation and give results on stability versus instability for
spherically symmetric static models. The paper ends with a conclusion in
section 4.

\section{Basic equations and their properties}

In this section we discuss the weak-field limit of Jordan, Brans-Dicke
theory \cite{2},\cite{20}. The JBD theory incorporates the Math principle,
which states that the phenomenon of inertia must arise from accelerations
with respect to the general mass distribution of the universe. Differences
between predictions of JBD theory and observing appeared at study of
deciding the field equation:

\begin{center}
\begin{equation}
R_{\mu \nu }-\frac 12Rg_{\mu \nu }=\frac{8\pi }{c^4\phi \ }T_{\mu \nu
}-\frac \omega {\phi ^2}\left( \phi _{,\mu }\phi _{,\nu }-\frac 12g_{\mu \nu
}\phi _{,\lambda }\phi ^{,\lambda }\right) -\frac 1\phi \left( \phi _{,\mu
;\nu }-g_{\mu \nu }\phi _{;\lambda }^{;\lambda }\right) ,  \label{1}
\end{equation}

\begin{equation}
\phi _{;\lambda }^{;\lambda }=-\frac{8\pi }{c^4\left( 3+2\omega \right) \ }T,
\label{2}
\end{equation}
\end{center}

Equations of gravitation field allow significant simplification, if velocity
of material point far less then velocity of light, so values of order v$^2$/c%
$^2$ possible neglect. In the case of weak field approximation value g$_{\mu
\nu }$ must extremely little differ from: 
\begin{eqnarray*}
g_{\mu \nu } &=&1\ \ for\ \mu =\nu =1,2,3,g_{00}=-1 \\
g_{\mu \nu } &=&0\ \ for\ \mu \neq \nu
\end{eqnarray*}

and squares of these deflections possible to neglect. Then

\begin{eqnarray}
\frac{\partial ^2x^\mu }{\partial t^2}=-c^2{\Gamma }_{00}^\mu ,  \label{2.2}
\end{eqnarray}

Moreover, in static case derived g$_{\mu \nu }$ on time possible to neglect
too. Then one can change $\Gamma _{00}^\mu \ $to $\Gamma _{\mu ,00}$, or -$%
\frac 12\frac{dg_{00}}{dx^\mu }\ $and equations of motion a material point (%
\ref{2.2}) for small velocities and weak field takes Newtonian form:

\[
\frac{\partial ^2x^\mu }{\partial t^2}=-\frac{\partial U}{\partial x^\mu } 
\]

where {\it U} is a gravitation potential, and

\begin{eqnarray}
g_{00}=1-\frac{2U}{c^2}.  \label{2.3}
\end{eqnarray}

From equations (\ref{1}), (\ref{2}) one can get

\begin{eqnarray}
R_{\mu \nu }+\frac \omega {\phi ^2}\phi _{;\mu }\phi _{;\nu }+\frac 1\phi
\phi _{;\mu ;\nu }=-\frac{8\pi }{c^4\phi \ }\left[ T_{\mu \nu }-\frac{%
1+\omega }{3+2\omega }Tg_{\mu \nu }\right] .  \label{2.4}
\end{eqnarray}

For the component 00 equations (\ref{2.4}) values of order v/c possible
neglect, except T$_{00}.\ $Component of energy momentum tensor T$_{00}=\rho
c^2$ consequently T = g$^{\mu \nu }T_{\mu \nu }=g^{00}T_{00}=-\rho c^2$, and

\[
R_{00}+\frac 1\phi \phi _{;0;0}=-\frac{8\pi }{c^2\phi }\frac{2+\omega }{%
3+2\omega }\rho 
\]

As far as derived on time and product $\Gamma _{\nu \lambda }^\mu \ $we
neglect, then

\[
R_{00}=\frac{\partial \Gamma _{00}^i}{\partial x} 
\]

since $\Gamma _{00}^i$ $\approx \Gamma _{i,00}\approx -\frac 12\frac{%
\partial g_{00}}{\partial x^i}$ then from (\ref{2.3})

\[
R_{00}=\frac 12\stackunder{i}{\sum }\frac{\partial ^2g_{_{00}}}{\partial
x_i^2}=\frac 12\triangle g_{00}=-\frac{\triangle U}{c^2}. 
\]

For scalar potential we have $\phi _{;0;0}=2\Gamma _{00}^i\frac{\partial
\phi }{\partial x^i}$ .

Finely from (\ref{1}) and (\ref{2})

\begin{eqnarray}
div\left( \phi \nabla U\right) =8\pi \frac{2+\omega }{3+2\omega }\rho ,
\label{2.5}
\end{eqnarray}

\begin{eqnarray}
\triangle \phi =-\frac{8\pi }{3+2\omega }\rho .  \label{2.6}
\end{eqnarray}

where $\rho $ is density of mater. Limiting transformation to Newton theory
of gravitation occurs when $\mid \omega \mid \ \rightarrow \infty \ $and $%
\phi $={\it const}.

\section{Dynamical stability}

We discuss the problem of dynamical stability of the equilibrium solutions
and consider small time dependent radial perturbations, which still preserve
spherical symmetry. Let's consider oscillations of configurations consisting
of ideal gas with adiabatic radial perturbations. Generally the solutions of
equations of stars oscillations are connected with big mathematical
difficulties, therefore we assume the approximation of small perturbation 
\cite{5}. Let's express {\it w}, {\it u} and {\it v} as radial, latitude and
meridianal components of speed in spherical coordinates:

\begin{eqnarray}
X=div\left( \frac{\partial \xi }{\partial t}\right) =\frac 1{r^2}\frac
\partial {\partial r}\left( r^2w\right) +\frac 1{r\ \sin \theta }\frac
\partial {\partial \theta }\left( u\ \sin \theta \right) +\frac 1{r\ \sin
\theta }\frac{\partial v}{\partial \varphi },  \label{2.7}
\end{eqnarray}

then from the equation of continuity

\begin{eqnarray}
\frac d{dt}\left( \rho _0+\rho \right) =-\left( \rho _0+\rho \right) X,
\label{2.8}
\end{eqnarray}

we have:

\begin{eqnarray}
\frac{\partial \rho }{\partial t}=-w\frac{\partial \rho _0}{\partial r}-\rho
_0X,  \label{2.9}
\end{eqnarray}

Throughout the paper, we use values with zero designate not perturbed
values, and without zero a perturbation of the value. In case of ideal gas
the adiabatic condition will be written down so

\begin{eqnarray}
\frac 1{P_0+P}\frac d{dt}\left( P_0+P\right) =\frac \gamma {\rho _0+\rho
}\frac d{dt}\left( \rho _0+\rho \right) .  \label{2.10}
\end{eqnarray}

In galactic dynamics $\gamma $ is less than 3 \cite{3} which means that no
polytropic stellar system can be homogeneous. In real stars $\gamma $ is the
variable and ranges from 1 to $\infty $, but possible changes of it are
rather small therefore we assume for simplicity, that it is a constant. We
use for convenience, values $\varepsilon $ and {\it g }determined as:

\begin{eqnarray*}
\varepsilon ^2 &=&\gamma \frac{P_0}{\rho _0}, \\
g &=&\nabla U.
\end{eqnarray*}

Then from an adiabatic condition we find

\begin{eqnarray}
\frac{\partial P}{\partial t}+w\frac{\partial P_0}{\partial r}=\varepsilon
^2\left( \frac{\partial \rho }{\partial t}+w\frac{\partial \rho _0}{\partial
r}\right) =-\rho _0\varepsilon ^2X.  \label{2.11}
\end{eqnarray}

From

\[
\frac{\partial P_0}{\partial r}=-g_0\rho _0 
\]

and (\ref{2.11}) we obtain

\begin{eqnarray}
\frac{\partial P}{\partial t}=\rho _0\left( g_0w-\varepsilon ^2X\right) .
\label{2.12}
\end{eqnarray}

The equations for gravitational potential satisfy

\[
div\left( \phi _0g_0\right) =8\pi \frac{2+\omega }{3+2\omega }\rho _0 
\]

Perturbation of a gravitational field is given by

\begin{eqnarray}
div\left( \phi g_0+\phi _0g\right) =8\pi \frac{2+\omega }{3+2\omega }\rho ,
\label{2.13}
\end{eqnarray}

The gravitational scalar consist the equilibrium and perturbed part which
satisfy to the equations

\begin{eqnarray}
div\left( \nabla \phi _0\right) =-\frac{8\pi \rho _0}{3+2\omega },
\label{2.14}
\end{eqnarray}
\begin{eqnarray}
div\left( \nabla \phi \right) =-\frac{8\pi \rho }{3+2\omega }.  \label{2.15}
\end{eqnarray}

The equations governing the linear perturbations are obtained by expanding
all functions to first order:

\begin{eqnarray}
\rho _0\frac{\partial u}{\partial t}=-\frac 1r\frac \partial {\partial
\theta }\left( P+\rho _0\varphi \right) ,  \label{2.16}
\end{eqnarray}

\begin{eqnarray}
\rho _0\frac{\partial v}{\partial t}=-\frac 1{r\ \sin \theta }\frac \partial
{\partial \varphi }\left( P+\rho _0\varphi \right) ,  \label{2.17}
\end{eqnarray}
\begin{eqnarray}
\rho _0\frac{\partial w}{\partial t}=-\frac{\partial P}{\partial r}-g_0\rho
-\rho _0g.  \label{2.18}
\end{eqnarray}

Furthermore, we suppose a time dependence of the form

\[
A\left( r,t\right) =A\left( r\right) e^{i\sigma t} 
\]

We differentiate (\ref{2.18}) on {\it t} and (\ref{2.12}) on {\it r} then
taking into account the equation of continuity it is received to

\begin{eqnarray}
\sigma ^2w-g_0^{\prime }w-g_0w^{\prime }-i\sigma g+\varepsilon ^2X^{\prime
}+g_0\left( 1-\gamma \right) X=0,  \label{3.19}
\end{eqnarray}

where the prime means derivative with respect to {\it r}. In a case of
radial pulsations (\ref{2.7}) it transformed to

\begin{eqnarray}
X=w^{\prime }+\frac{2w}r,  \label{3.20}
\end{eqnarray}

The equations (\ref{2.13}) and (\ref{2.15}) transforms to

\begin{eqnarray}
{i\sigma g}=-\frac{8\pi \left( 2+\omega \right) \rho _0w}{\left( 3+2\omega
\right) \phi _0}-\frac{i\sigma g_0\phi }{\phi _0},  \label{2.20}
\end{eqnarray}

\begin{eqnarray}
{i\sigma }\frac{\partial \phi }{\partial r}=\frac{8\pi \rho _0}{3+2\omega }w.
\label{2.21}
\end{eqnarray}

Then the equation (\ref{3.19}) takes a form

\begin{eqnarray}
-\sigma ^2w &=&-g_0w^{\prime }+\frac{i\sigma g_0\phi }{\phi _0}+\varepsilon
^2\left( w^{\prime \prime }+\frac{2w^{\prime }}r-\frac{2w}{r^2}\right) + 
\nonumber  \label{u2.22} \\
&&+g_0\left( 1-\gamma \right) \left( w^{\prime }+\frac{2w}r\right) +\frac
2rwg_0+\frac{g_0\phi _0^{\prime }w}{\phi _0}
\end{eqnarray}

Performing a change of variable defined by {\it w} = {\it r f }({\it r})
than from (\ref{u2.22}) we obtain for {\it f}

\begin{equation}
\varepsilon ^2f^{\prime \prime }+f^{\prime }\left( \frac{4\varepsilon ^2}%
r-\gamma g_0\right) +f\ \left[ \sigma ^2+g_0\left( \frac{4-3\gamma }r+\frac{%
\phi _0^{\prime }}{\phi _0}\right) \right] +\frac{g_0}{r\ \phi _0}\int \frac{%
8\pi \ r\ f\ \rho _0}{3+2\omega }dr=0,  \label{u2.23}
\end{equation}

the pulsation equation transforms to the following Eddington equation in the 
$\ $limit $\omega \rightarrow \infty \ $and $\phi ={\it const}:$

\begin{equation}
\varepsilon ^2f^{\prime \prime }+f^{\prime }\left( \frac{4\varepsilon ^2}%
r-\gamma g_0\right) +f\left[ \sigma ^2+\left( 4-3\gamma \right) \frac{g_0}%
r\right] =0.  \label{2.24}
\end{equation}

Let's put as a first approximation {\it f} equal to a constant then from (%
\ref{u2.23})

\[
\sigma ^2=g_0\left( \frac{4-3\gamma }r+\frac{8\pi \ \int r\ \rho _0dr}{%
r\left( 3+2\omega \right) \phi _0}+\frac{\phi _0^{\prime }}{\phi _0}\right)
. 
\]

The stability condition one can obtain when put $\sigma ^2$ = 0, in this
case critical value for $\gamma $ is

\begin{eqnarray}
\gamma _{cr}=\frac 43+\frac{8\pi \ \int r\ \rho _0dr}{3\left( 3+2\omega
\right) \phi _0}+\frac{r\ \phi _0^{\prime }}{3\phi _0}.  \label{uu2.25}
\end{eqnarray}

To begin to understand this problem we have model of gas density
distribution for clusters of galaxies \cite{3}, with functions of a matter
distributions for spherically symmetric objects:

\begin{eqnarray}
\rho _0=\rho _c\left( 1+\left( \frac rR\right) ^2\right) ^{-1.5\beta },
\label{uu2.26}
\end{eqnarray}

where the core radius R, central density $\rho _c$ , and the number are
parameters varies from cluster to cluster but has typical value the order of
2/3 which implies that the gas mass generally increases linearly with
radius. There are analogous form of density distribution for globular star
cluster \cite{7}, in this case $\beta $ = 1. Comparing this relation with a
numerical solution of Lane-Emden equation for an isothermal sphere \cite{c33}
one can say that for particular case the relative error is less than 5 \%.
Although these models are simplistic, it exhibits many of the key features
of the more complex problem. Using relations describing the clusters of
galaxies like (\ref{uu2.26}), it is possible to give a qualitative answer to
the question that reaches critical values of $\gamma $.

Thus, under the assumption of matter distributions (\ref{uu2.26}) we have
find from (\ref{2.14}) expression for not perturbed value of scalar field $%
\phi _0$

\begin{eqnarray}
\phi _0=\int \frac{-8\pi \int \rho _c(1+\frac{r^2}{R^2})^{-\frac{3\beta }%
2}r^2dr+(3+2\omega )C_1}{(3+2\omega )r^2}dr+C_2,  \label{uu2.27}
\end{eqnarray}

In the previous expressions one can exlain the $\phi _0\ $using the Gauss
hypergeometric function.

For large radius where the effect of the central conditions is very weak the
solution should asymptotycally approach the exterior solution. It is know
that the scalar field is a constant outside the matter distributions \cite
{c8}. In empty space there is usual Newtonian universe, but inside
''gravitation constant'' depend on matter distributions. Thus, the
integration constants C$_1$ and C$_2$ are determined by matching the
interior solution (\ref{uu2.27}) to the usual exterior Newtonian vacuum
solution $\phi _0$={\it G}, $\phi _0^{\prime }$=0.

To construct a spherically symmetric configurations for given matter
distributions (\ref{uu2.26}), we choose $\beta $ = 1 and $\beta $ = 3/2.
Knowing $\phi _0$ and $\rho _0$ allows for the determination of the critical
values of $\gamma \ $for linear adiabatic radial perturbations of
spherically symmetric gas spheres, using equation (\ref{uu2.25}). The whole
structure of the perturbated gas spheres is thus determined.

The region of classic Newtonian stability is the rectangle 0 \TEXTsymbol{<} 
{\it n }\TEXTsymbol{<} 3. However, the conjecture that $\gamma $ \TEXTsymbol{%
>} 4/3 is a necessary and sufficient condition for stability in the
Newtonian approximation of Jordan-Brans-Dicke theory of gravitation is shown
to be false. The critical values of $\gamma \ $depends on the values of the
Brans-Dicke parameter $\omega $ and on the values of the central dencity $%
\rho _c$. For both casses here $\beta $ = 1 and $\beta $ = 3/2 for negative $%
\omega $ is seen $\gamma \ $ to be less than 4/3 in the instability region.
Since scalar field effects are stabilizing, one expects stability in wider
area than of the Newtonian region.

\section{Conclusion}

In the present work, we have thoroughly analyzed static spherically
symmetric configurations in the framework of the Newtonian approximation of
Jordan-Brans-Dicke theory of gravitation. The classic result for adiabatic
radial pulsations of Newtonian gas spheres is given by the marginal
stability $\gamma $ \TEXTsymbol{>} $\frac 43$. Since scalar field effects
are stabilizing, one expects stability in extended area of the Newtonian
region. The behavior of the stable equilibrium solutions for the spherically
symmetric configurations considered here, it emerges that the more compact a
model is, the more stable it is. Furthermore, in the Jordan, Brans-Dicke
theory exist the stable configurations for any polytropic index. The similar
effect in classic Newtonian theories does not exist. Whichever it might be,
it is very likely that the same phenomena could also occur for clusters of
galaxies, such as neutron stars and white dwarfs. In this sense, the results
obtained in this paper can be regarded as of a general nature.

The goal of the present work was to present a stability analyze for model of
gas density distribution. This configurations can be used to compare with a
numerical stability code for configurations with polytropic equations of
state. Together with this, a number of new physical features have been
displayed concerning the radius-mass relation.


\begin{thebibliography}{99}
\bibitem{1}  C.H. Brans, {\it Gravity and the Tenacious Scalar Field},
gr-qc/9705069, (1997).

\bibitem{10}  C. Will, {\it The Confrontation between General Relativity and
Experiment, }gr- qc/ 0103036 (2001).

\bibitem{2}  P.Jordan , ''Schwerkraft und Weltall'', Braunshweig, (1955).

\bibitem{20}  C. Brans and R. H. Dicke, Phys. Rev. {\bf 124} (1961) 925.

\bibitem{3}  J. Binney, S.Tremaine, Galactic Dynamics, Princeton University
Press, (1987)

\bibitem{c31}  S.A.Bludman, Ap. J. 183, 637-648, (1973).

\bibitem{c32}  E.N.Glass, A. Harpaz, Mon. Not. R. astr. Soc. 202, 159-171,
(1983).

\bibitem{c33}  F.K. Liu,{\it \ Polytropic gas spheres: An approximate
analytic solution of the Lane-Emden equation, }astro- ph/ 9512061 (1995).

\bibitem{c34}  V. Bashkov, Yu. V. Davydova, P.V. Lopukhov, Exact solution of
Lane-Emden equation for equilibrium of Newtonian stars, ''Gemetrization of
Physic III'', Kazan, (1997)

\bibitem{c35}  P. M. Sa, {\it Polytropic stars in three-dimensional
spacetime, }gr-qc/0302074, (2003).

\bibitem{4}  R. Kippenhahn, A. Weigert ''Stellar Structure and Evolution'',
Berlin, (1990), Springer-Verlag

\bibitem{5}  S.Chandrasekhar, Ap. J. 130, 664-674, (1964).

\bibitem{7}  I.R.King, The structure of star cluster, Astron. J., 71, 64,
(1966).

\bibitem{c8}  S. Kozyrev, {\it Properties of the static, spherically
symmetric solutions in the Jordan, Brans-Dicke theory}, gr-qc/0207039 (2002).
\end{thebibliography}
\end{document}